\documentclass[prl,aps,twocolumn,showpacs]{revtex4}
\usepackage{amsmath,graphicx,epsfig}
\def\fig_width{8.6 cm} 

\newlength{\defbaselineskip}
\newcommand{\setlinespacing}[1]%
           {\setlength{\baselineskip}{#1 \defbaselineskip}}

\begin{document}

\title{A Three Dimensional Lattice of Ion Traps} 

\author{K. Ravi,$^1$ Seunghyun Lee,$^1$ Arijit Sharma,$^1$ Tridib Ray,$^1$ G. Werth$^2$ and S. A. Rangwala$^1$}
\email{sarangwala@rri.res.in}
\affiliation{$^1${Raman Research Institute, Sadashivanagar, Bangalore 560080, India}\\
$^2${Institut f\"ur Physik, Johannes-Gutenberg-Universit\"at, D-55099 Mainz, Germany}}
\date{\today}


\begin{abstract}

We propose an ion trap configuration such that individual traps can be stacked together in a three dimensional simple cubic arrangement. The isolated trap as well as the extended array of ion traps are characterized for different locations in the lattice, illustrating the robustness of the lattice of traps concept. Ease in the addressing of ions at each lattice site, individually or simultaneously, makes this system naturally suitable for a number of experiments. Application of this trap to precision spectroscopy, quantum information processing and the study of few particle interacting system are discussed. 

\end{abstract}
\pacs{32.80.Pj, 29.30.Ep, 95.55.Sh, 03.67.-a, 21.45.+v}







\maketitle


The trapping of ions with static~\cite{Deh90} and time varying~\cite{Pau90} fields has been central to many important developments in physics and chemistry for the past 50 years. Ion traps have been applied to precision spectroscopy~\cite{Deh90}, precision mass determination~\cite{Bla06}, mass analysis~\cite{Mar05}, atomic ion clocks~\cite{Ros08} and most recently in quantum information processing~\cite{Bla08}. Ion trap experiments most typically focus on the ability to trap (1) a single ultra-cold ion for extended interrogation, (2) a collection of ions in the form of a spatially extended cloud~\cite{Maj04} or (3) an ordered crystal of ions~\cite{Die87,Win87}. Multiple species of ions can also be trapped simultaneously~\cite{Mol00}. Common to all of these experiments is the study of ion(s) within isolated ion traps. Recently experiments have been performed to communicate between trapped ions in remote experiments~\cite{Rie04}. Different trap configurations have been developed to meet specific requirements, starting from conventional 3-dimensional (3-D) traps with hyperbolic electrodes~\cite{Pau90} to cylindrical traps~\cite{Gab84}, linear traps~\cite{Dre00} and various kinds of planar traps~\cite{Hen06,Kie02,Sei06,Sta05}.

In this Letter we propose a novel configuration for ion trapping which is three dimensionally symmetric by its geometry. It resembles the configuration used by Wuerker {\it et al.}~\cite{Wue59} in 1959. An important consequence of the present structure is its suitability for the construction of several compact, stacked, virtually independent ion traps in the the same apparatus. Here we discuss the fundamental structure underlying the proposed trap and its operating scheme. Parameters for both the isolated fundamental trap and the system of stacked traps are determined and compared, establishing the robustness of the configuration. The motion of ions in the traps is characterized, building a strong case for several experimental applications with this trapping scheme, some of which are sketched to illustrate its versatility. 

\begin{figure}
\includegraphics[width=3.00 cm]{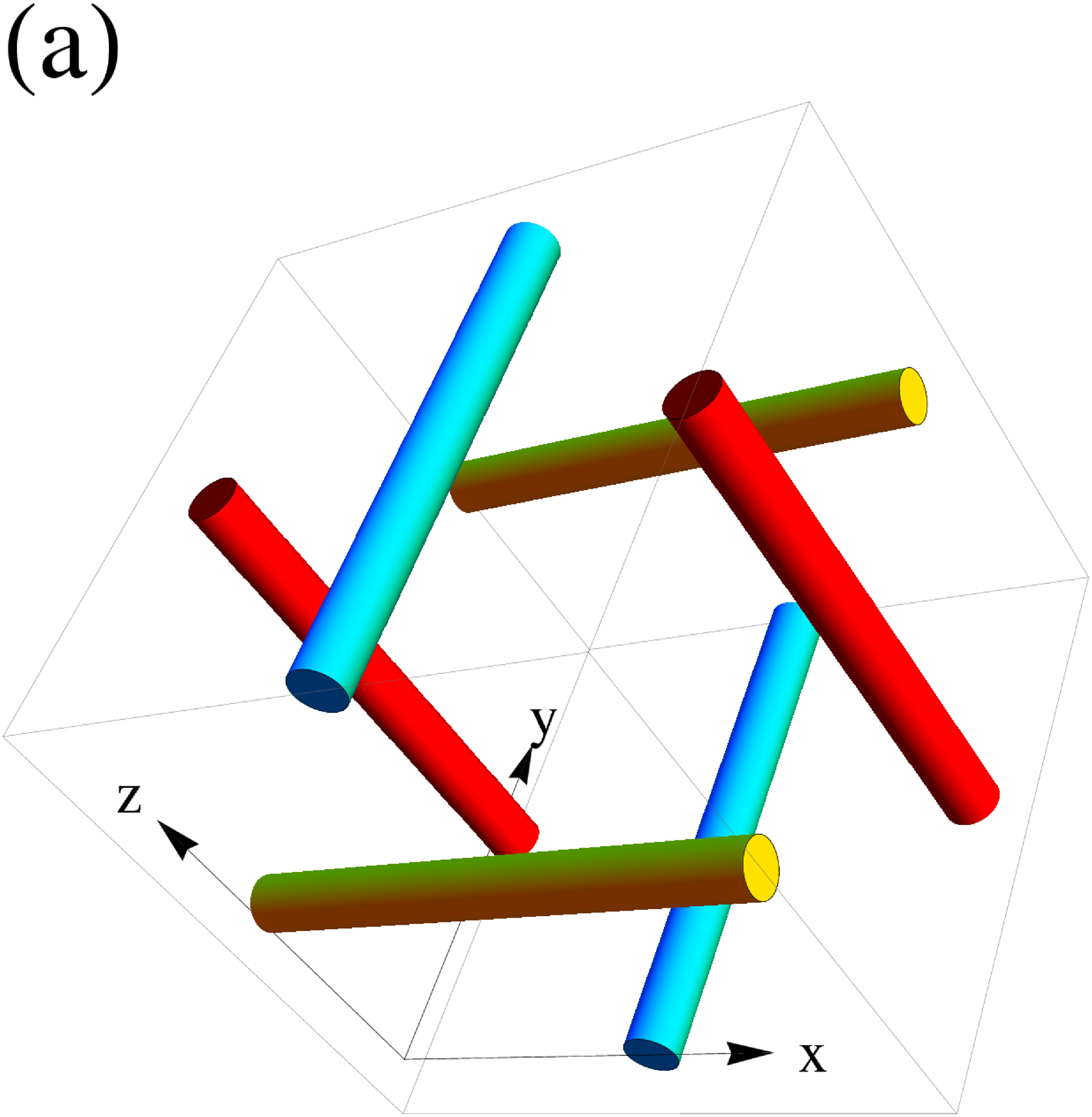}
\includegraphics[width=4.25 cm]{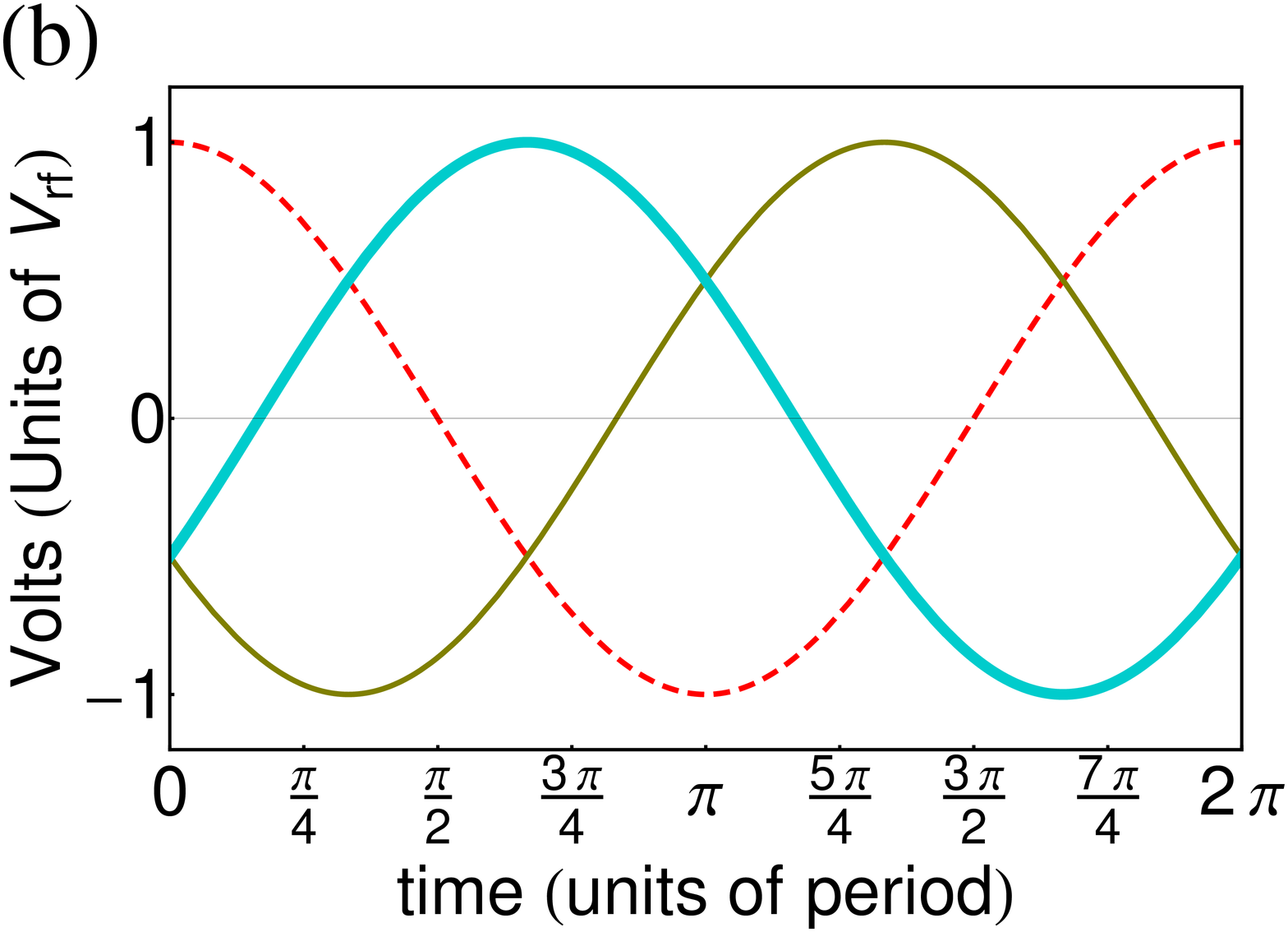}
\includegraphics[width=4.25 cm]{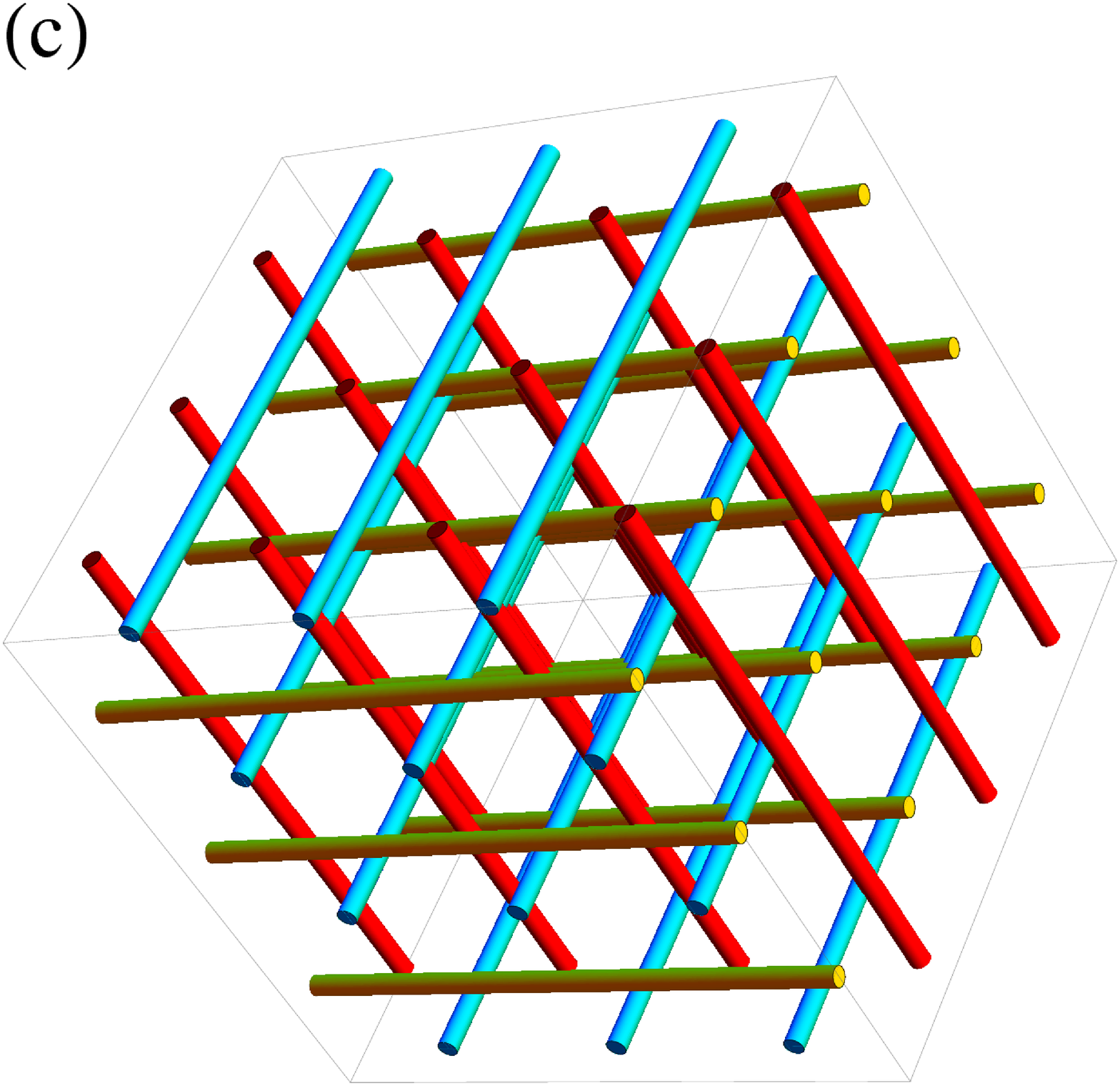}
\includegraphics[width=3.00 cm]{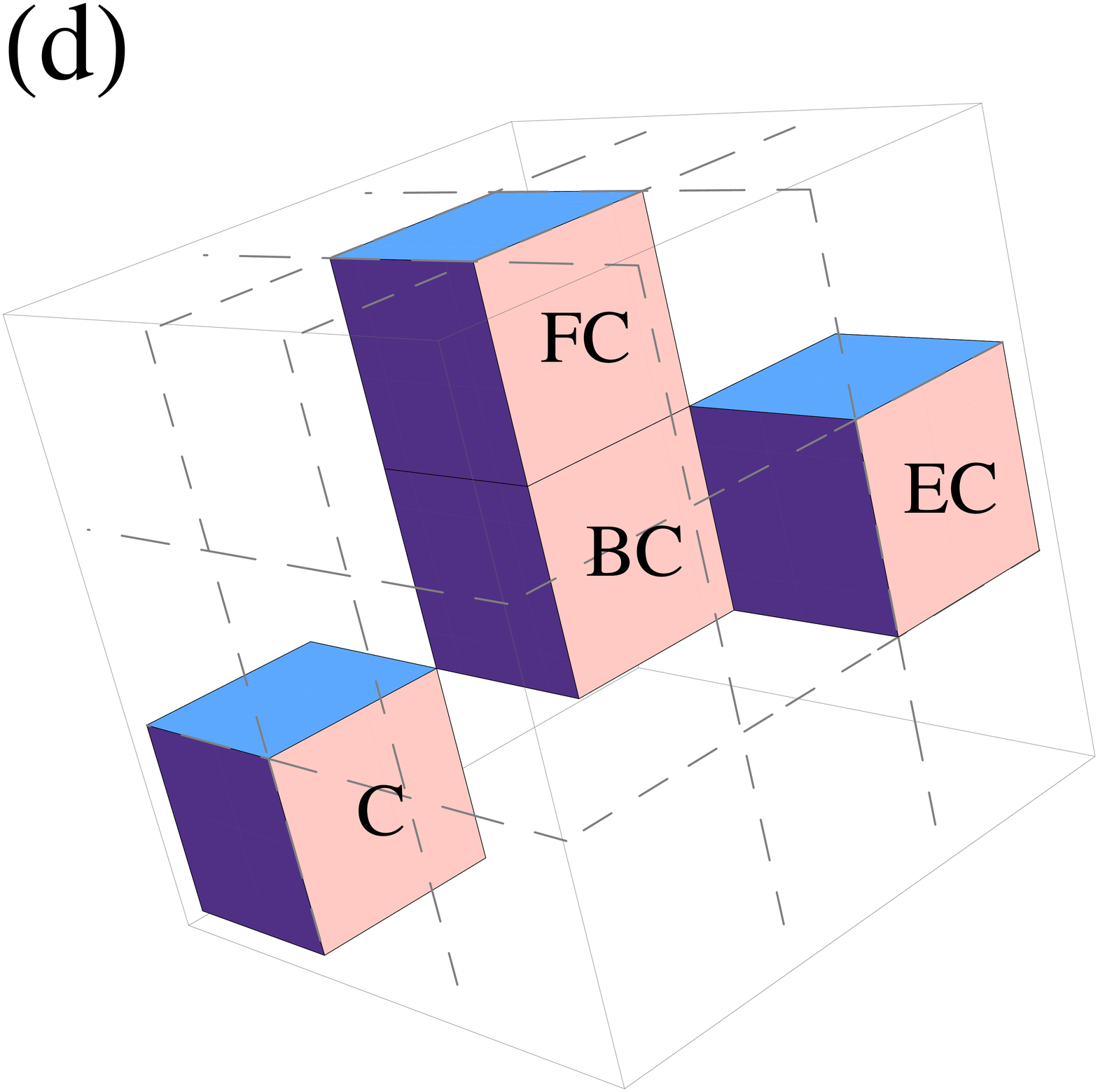} 
\caption{(color online) Structure and operation of the 3-D ion trap. Part (a) shows the basic electrode configuration of the ion trap. Three pairs of parallel electrodes are arranged along the cartesian axes and are electrically contacted. The relative phase difference of $2\pi/3$ between the orthogonal electrode sets is shown in (b), where (V$_x$,V$_y$,V$_z$) are represented by (thin, thick, dashed) lines respectively. The extension of the trap in (a) to 27 traps in simple cubic arrangement is illustrated in (c). 4 distinct types of coordination symmetries for traps at different locations emerge (d). Representative cells of different coordination are identified as BC(6), FC(5), EC(4) and C(3), where the bracketed number is the location dependent nearest neighbor cells.}
\label{Fig:3DLatticeConfig}
\end{figure}

The fundamental building block for the Lattice of Ion Traps (LIT) is illustrated in Fig.~\ref{Fig:3DLatticeConfig}(a). The trap is constructed with 3 parallel pairs of cylindrical electrodes along the cartesian axes. When extended this creates compact, well isolated individual traps with large optical access, permitting optical communication between different traps and ease of construction. Typical dimensions could be electrode diameter 0.5mm and a center-to-center separation of 5mm. These are the dimensions used in our simulations. The dimensions can be changed without qualitatively changing the characterization below. 

Time varying potentials create the 3-D ion trap, when applied to the equipotential, parallel electrodes in Fig.~\ref{Fig:3DLatticeConfig}(a). The phase relation for the amplitude of the applied radio frequency (RF) field between electrode sets along $\hat{x}$, $\hat{y}$ and $\hat{z}$ is illustrated in Fig.~\ref{Fig:3DLatticeConfig}(b). This field configuration creates the rotating saddle potential which traps the ions in a 3-D symmetric configuration over one full cycle of the RF. This symmetry lends itself to extension, to form the LIT illustrated in Fig.~\ref{Fig:3DLatticeConfig}(c). In general any  $l\times m\times n$ number of equivalent traps can be constructed by building an appropriate number of electrodes. The RF phase relation remains identical to the fundamental trap for the extended sets of parallel electrodes. Fig.~\ref{Fig:3DLatticeConfig}(c) illustrates the $3\times 3\times 3$ version of the LIT as it is the lowest order cubic arrangement that provides lattice sites with different near neighbor coordination. The four distinct locations of ion traps in this arrangement are Body Center (BC), Face Center (FC), Edge Center (EC) and Corner (C). We show below that the site specific effects on the trapped ions in individual traps are too small to be significant.

The ion trap operation is simulated for the $^{40}$Ca$^+$ ion. The potentials for the ion trap(s) are generated using SIMION. The equations of motion in the time varying potentials are solved numerically using Mathematica. The rotating saddle potential results in ion motion with a macromotion frequency $\nu_m$ whose amplitude is modulated by the driving frequency of the trap, $\nu_{rf}$. Since $\nu_m$ can be significantly less than $\nu_{rf}$ care was taken in solving for the trajectory of the ion in three dimensions. Characterization simulations for the isolated ion trap and the LIT potential were done by solving for the full $3\times 3\times 3$ cell configuration. The isolated trap solutions are then compared with the corresponding solutions within the individual traps at the BC, FC, EC and C locations. 

Operationally, apart from the dimensions of the trap, 3 parameters characterize the symmetric 3-D ion trap. These are frequency of operation, amplitude of the RF and the trap depth for the ion. Equal amplitudes of the RF voltage V$_{rf}$  applied as in Fig.~\ref{Fig:3DLatticeConfig}(b) yields $\nu_{m;x}=\nu_{m;y}=\nu_{m;z}$. No dc field is applied to the electrode in our trap configuration. As a consequence of this the standard Paul trap representation in terms of stability parameters $a$ and $q$, related to the Mathieu differential equation, which governs the motion of ions in a RF trap~\cite{Pau90} is not relevant as $a = 0$. However $$q = 4 Q V_{rf}/{M (2 \pi \nu_{rf})^2 R_0^2},$$ where $Q$ represents the charge state of the ion, $M$ the mass and $R_0=0.0025$m the distance of the electrode from the trap center i.e. the size of the trap, remains a good expression for trap characterization.

\begin{figure}
\center
\includegraphics[width=7.5 cm]{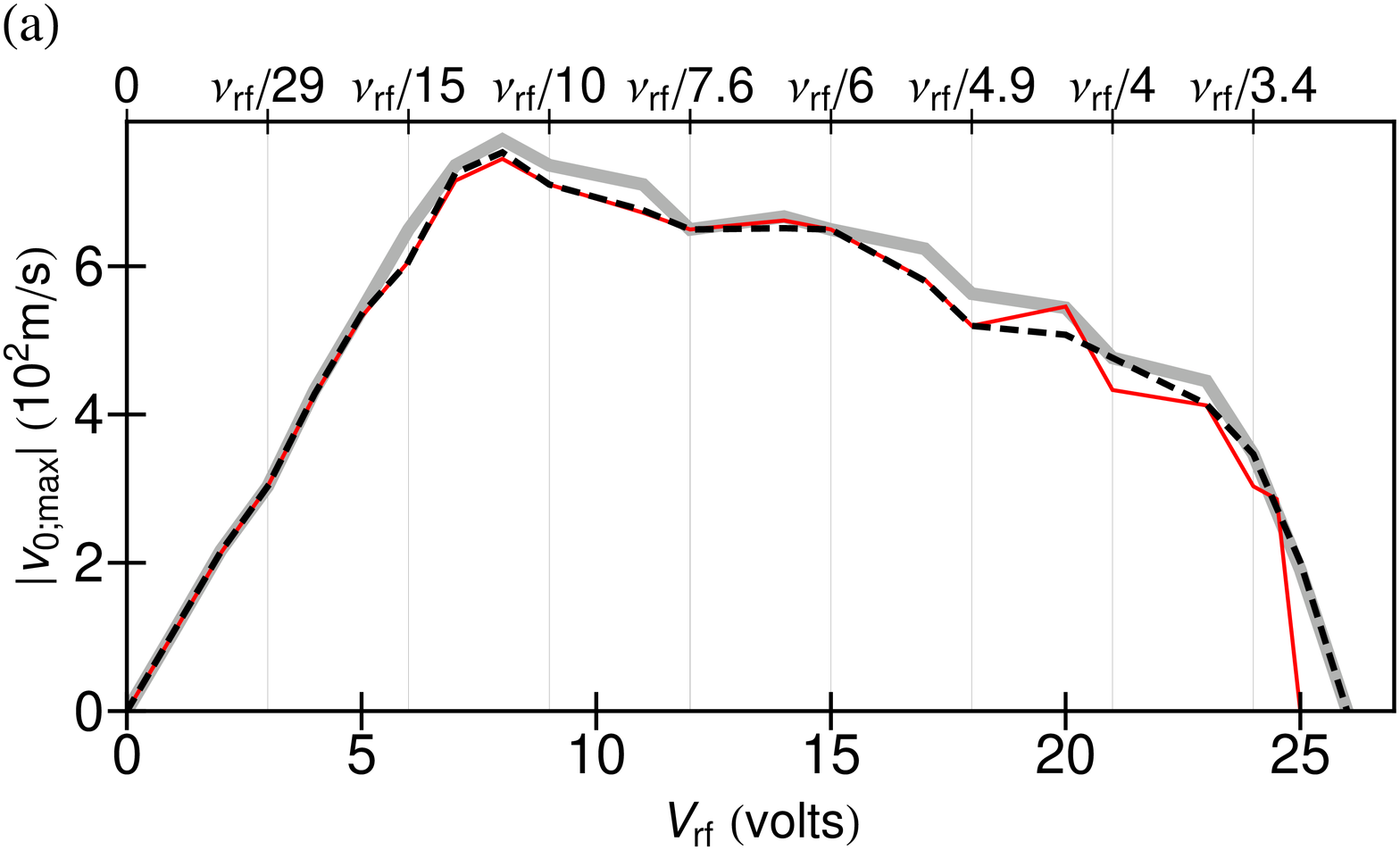} 
\includegraphics[width=3.75 cm]{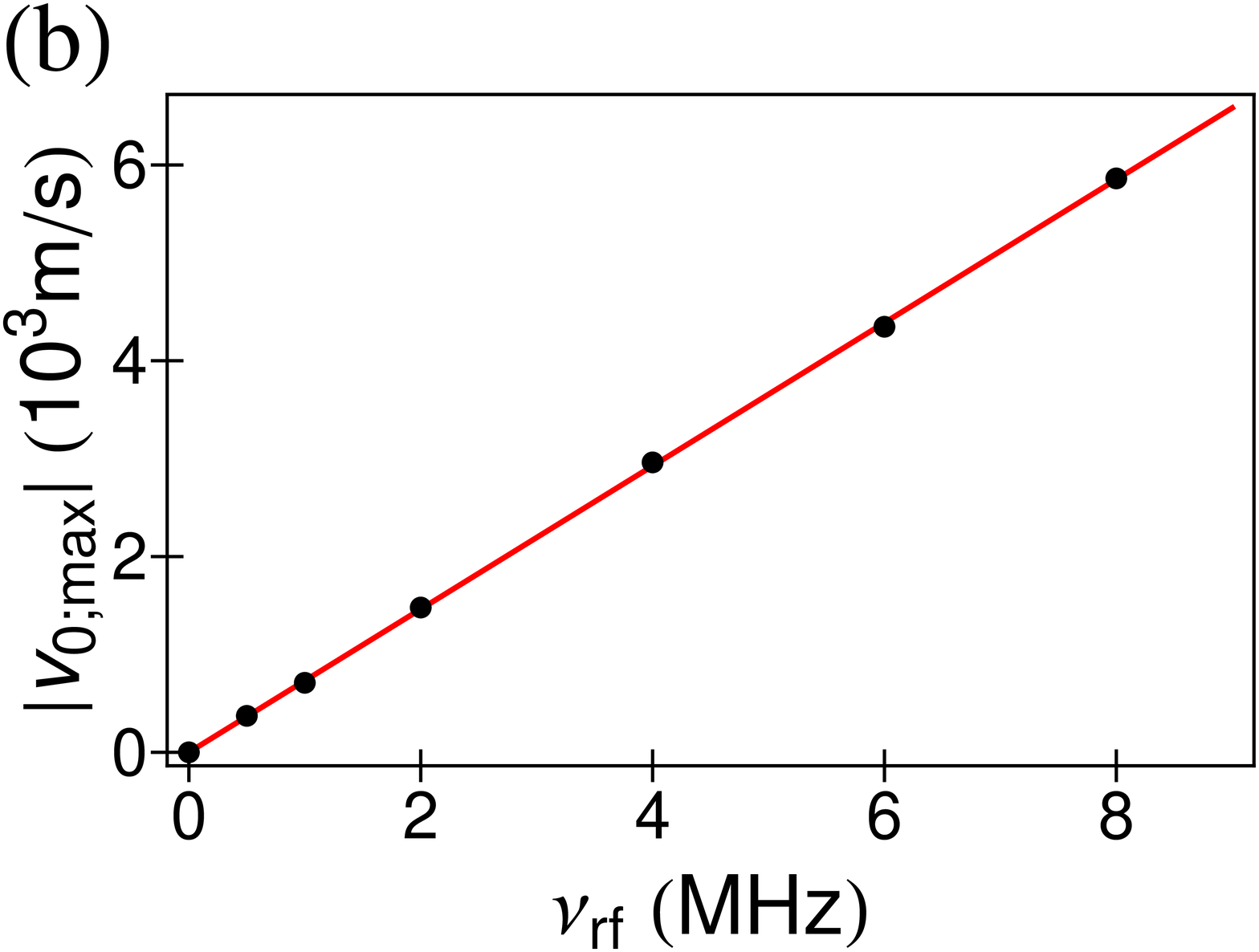} 
\includegraphics[width=3.75 cm]{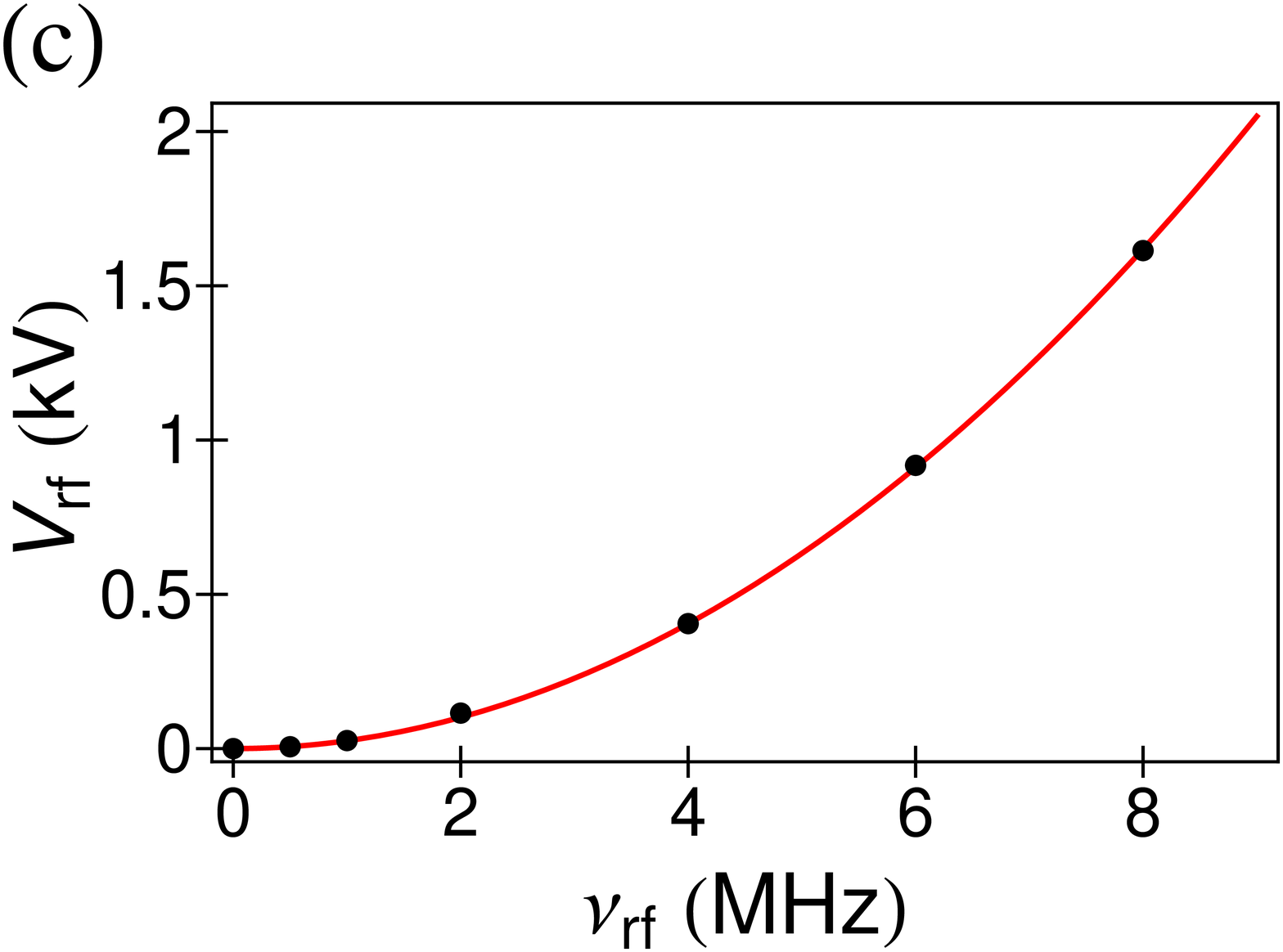}
\caption{(color online) Details of trap characterization. Part (a) illustrates the region of stability in the $\left|{\bf v}_0\right|$ vs. V$_{rf}$ plane at $\nu_{rf}=$ 1MHz. The region contained within the the curve is the stable region. The dashed(black), thin(red) and thick(grey) lines are the stability bounds for the single trap of Fig.~\ref{Fig:3DLatticeConfig}(a), the most configurationally distinct BC and C cells in the LIT respectively. Part (b) plots the maximum initial velocity that is trapped $\left|{\bf v}_{0;max}\right|$ and (c) the maximum amplitude of the RF field which traps the $^{40}$Ca$^+$ ion as a function of $\nu_{rf}$. The line is the least square fit to the points, for (b) $\alpha \nu_{rf}$ and $\beta \nu^{2}_{rf}$ for (c).}
\label{Fig:TrapCharacteristics}
\end{figure}

The nature of the present trap is characterized by relating the RF amplitude V$_{rf}$, the magnitude of the initial velocity of the trapped ion $\left|{\bf v}_0\right|$ and $\nu_{rf}$. Fig.~\ref{Fig:TrapCharacteristics}(a) illustrates the boundary of the region of stability in the $\left|{\bf v}_0\right|-$V$_{rf}$ plane, where $\nu_{rf}=$ 1MHz, for the single trap, BC and C locations. The initial position of the ion is at the trap center and initial velocity is $\bf{v}_0$ with equal components in the 3 orthogonal directions. The small variations in the trapping region in Fig.~\ref{Fig:TrapCharacteristics}(a) arise due to trapping potential at the edge of the physical trap. A barely trapped ion samples the outer reaches of the individual trap potential where, in addition to intrinsic anharmonicity, the trap at its edges, is susceptible to its immediate neighborhood. Thus the differences due to the trap location in the LIT are minor. The top axis of Fig.~\ref{Fig:TrapCharacteristics}(a) marks out the macromotion frequency $\nu_m$ as a fraction of $\nu_{rf}$, for the isolated trap and is only dependent on the magnitude of V$_{rf}$. As V$_{rf}$ increases the amplitude of the micromotion increases rapidly, thus taking the ions to the trap boundary and ejecting them. Deeper trap potentials therefore do not imply higher initial kinetic energies of trappable ions, as illustrated by the fall in the $\left|{\bf v}_{0;max}\right|$ for higher V$_{rf}$ values. Thus the suitable operating point of the ion trap is  the initial region of linear response of $\left|{\bf v}_{0;max}\right|$ to V$_{rf}$ with $q \leq 0.29$ for V$_{rf}\leq 8$V at $\nu_{rf}=$ 1MHz. Generating multiple $\left|{\bf v}_0\right|-$V$_{rf}$ stability plots for different $\nu_{rf}$ and compiling the results provides the data for Fig.~\ref{Fig:TrapCharacteristics}(b) and (c). Fig.~\ref{Fig:TrapCharacteristics}(b) illustrates that $\left|{\bf v}_{0;max}\right|$ scales as $\nu_{rf}$, quantifying the maximum trappable initial velocity of the ion. Fig.~\ref{Fig:TrapCharacteristics}(c) demonstrates that the maximum trapping voltage V$_{rf;max}$ scales as $\nu^2_{rf}$ consistent with the behavior for constant $q$, here $q = 0.908$, which barely traps the zero velocity ion. Thus the bounds for the operation of the present ion trap are well characterized via the scaling functions shown. 

\begin{figure}
\center
\includegraphics[width=7.5 cm]{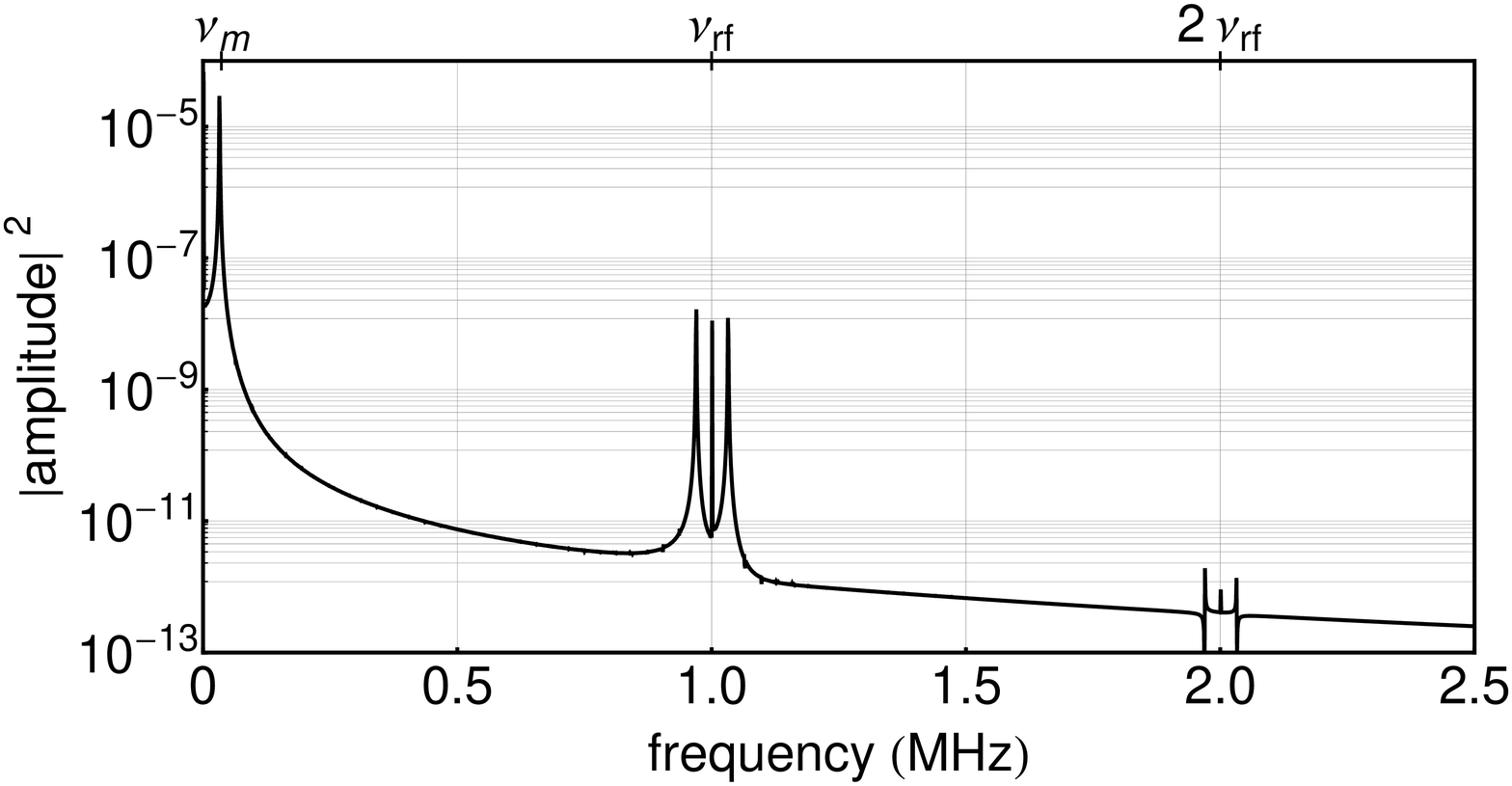} 
\caption{Power spectrum of the $\hat{x}$ component of the trajectory for a low $\left|{\bf v}_0\right|$ ion computed at $\nu_{rf} = 1$MHz, V$_{rf} = 3$V. A strong macromotion peak at $\nu_m$ and significantly weaker peaks at  $\nu_{rf}$, $\nu_{rf}\pm \nu_m$, $2 \nu_{rf}$ and $2 \nu_{rf}\pm \nu_m$ are seen.}
\label{Fig:ReprFTPowerSpectrum}
\end{figure}

Symmetric operation of the trap gives identical macromotion frequency $\nu_m$, in the orthogonal $\hat{x}$, $\hat{y}$ and $\hat{z}$ direction. The micromotion of the trajectory components is synchronous with the drive field frequency, $\nu_{rf}$. At the trap center and for a large range of $\left|{\bf v}_0\right|$, $\nu_m$ is unchanged and the power spectrum Fig.~\ref{Fig:ReprFTPowerSpectrum} remains free of sidebands, indicating that the trap is harmonic. Due to the large difference between $\nu_m$ and $\nu_{rf}$ in the $q \leq 0.29$ range of operation, it is reasonable to separate the ion motion into 2 components and the slow motion at $\nu_m$ can be considered separate, while averaging over the fast oscillation at $\nu_{rf}$. This is indeed exhibited clearly in the amplitude of the power spectrum of the fourier transform along the representative $\hat{x}$ direction in Fig.~\ref{Fig:ReprFTPowerSpectrum}. In the small $\bf{v}_0$ and low $q$ region of operation, $\nu_m$ is the same across traps with different coordination numbers.

The error in $\nu_m$ determination from our simulations is $< \pm 0.01\nu_{rf}$. The likely source of this error is the numerical solution to the Laplace equation, which at our computational convergence limit is marginally different from the true value of the potential. The complexity of the electrode structure here does not allow a fruitful comparison for the motion of a trapped ion in the computed trap and its idealised analytic form.

\begin{figure}
\center
\includegraphics[width=7.5 cm]{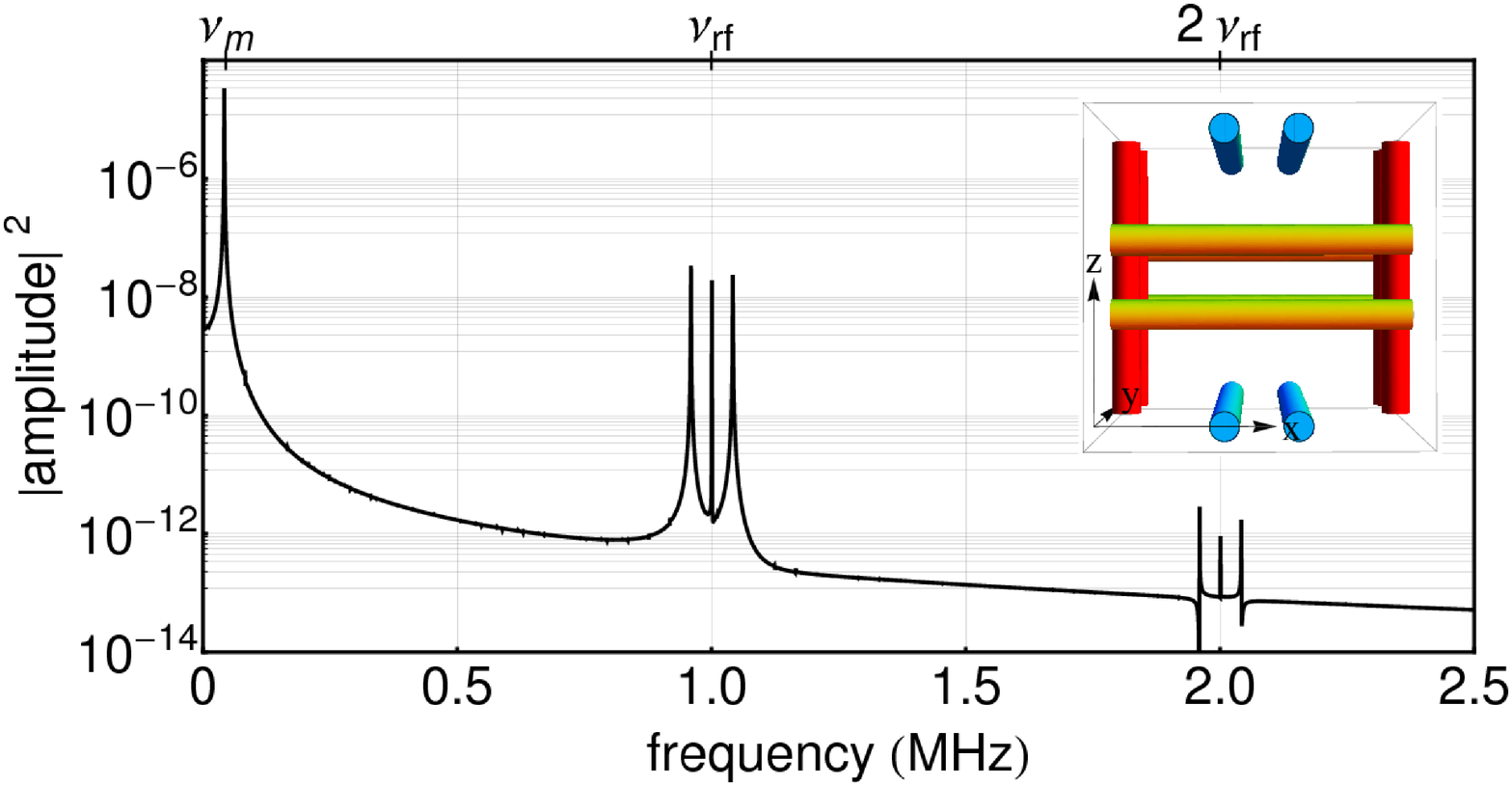} 
\caption{(color online). Power spectrum of the $\hat{x}$ component of the trajectory for a low $\left|{\bf v}_0\right|$ ion computed at $\nu_{rf} = 1$MHz, V$_{rf} = 3$V for the ion trap with split electrodes (inset). The resulting spectrum is qualitatively similar to that of Fig.~\ref{Fig:ReprFTPowerSpectrum}.}
\label{Fig:SplitElectrodes}
\end{figure}

Structurally the electrodes of the trap illustrated in Fig.~\ref{Fig:3DLatticeConfig}(a) and (c) do not allow a face on view of the ion. This can be easily circumvented by splitting each of the six electrodes into two closely spaced electrodes as shown in Fig.~\ref{Fig:SplitElectrodes}(a). The resulting trap is similar in performance to the trap in Fig.~\ref{Fig:3DLatticeConfig}(a) as illustrated in Fig.~\ref{Fig:SplitElectrodes}(b) by the power spectrum of the 
$\hat{x}$ component of the ion trajectory. 

Loading the LIT with the desired number of ions is most efficiently achieved by resonance enhanced two photon ionization. Briefly, the RF fields could be applied while a vapor of the parent neutral floods the lattice. Application of the two laser frequencies participating in the ionization, along two distinct directions of easy viewing through the lattice would create ions only at the beam intersection. Thus ion(s) can be loaded in a controlled way in each ion trap cell, close to the center of the individual traps. Laser cooling of ions to ultracold temperatures, now well established, is easily done along the easy viewing directions. Below we explore some applications which exploit the symmetry and multiplicity of the ion trap.

The most precise spectroscopic measurement on a system would be on an isolated, trapped single particle, nearly at rest. However in the specific case of a single ion confined within an ion trap, a serious constraint is presented by the low flux of the photons and hence the signal/noise (S/N) in the detection. Adding more than one ion per ion trap leaves the system vulnerable to perturbations that could compromise the precision of the measured transition frequencies. With the experimental setup proposed here, we have the ability to preserve the isolation of the individual ions in each trap, while increasing the detected fluorescence rate. 

Specifically the performance of single ion optical clocks may be significantly improved when a single ultra-cold ion is trapped within each cell of the LIT. The stability of clocks expressed by the Allen variance $$\sigma=({\Delta \nu}/{\nu}) (S/N)^{-1/2} \times \tau^{-1/2},$$ where $\nu$ is the transition frequency and $\Delta \nu$ the observed linewidth and $\tau$ is the integration time. In the arrangement illustrated in Fig.~\ref{Fig:3DLatticeConfig}(c), 27 such trapped ions can be interrogated simultaneously, increasing the $S/N$. This results in a decrease in the short term instability by about a factor of 5. 

Information processing with ion traps allows the most precise control over qubits thus far. However one practical limitation faced by most ion trap quantum information experiments is scalability and to an extent, precise individual addressing of the ions trapped in the ground vibrational state of a trap. Within the LIT individual cell traps can be addressed with ease due to the relatively large separations between neighboring cells. Also, for low  $\bf{v}_0$ and low $q$ operation the trap frequency $\nu_m$ is coordination number independent. In addition traps with the same coordination number are by definition identical. Thus if one desires to operate with only one qubit per site, this system is ideally suited. 

Further the ground state of the effective harmonic oscillator trap for the 3-D ion trap configuration is non-degenerate. However for the center of mass oscillations, the excited states in the symmetric operation mode of the trap exhibit degeneracy. Thus the usual cooling mechanisms (laser and sideband) essential for the population of the ground vibrational state of the trap with the ion(s) are unchanged. In addition, specific numbers of ions can be loaded per site in the 3-D oscillator ground state. In the case of 2 ions per trap cell, proceeding with entangling them with available protocols using $\pi/2$ and $\pi$ pulses with appropriate detunings, or performing gate operations~\cite{Cir95} is feasible. The one difference which arises due to the high symmetry is an overall rotational motion about the center of mass of the trapped ions which, due to the distance between two ions in the trap ground state (several $\mu$m) is very slow ($\approx 1$Hz) compared with other relevant time constants in such systems. Thus entanglement of ions within a specific cell is entirely possible with present day technology. Applications to studies of entanglement transfer, swapping~\cite{Rie08} and manipulation would be greatly facilitated by the presence of the lattice. The fact that by asymmetric application of either constant or RF voltage the trap degeneracy can be lifted for the excited states offers other degrees of freedom for exploitation in the future. Thus the LIT offers a new architecture for quantum information processing with ions~\cite{Kie02}. 

Another rich area of study with the three dimensional ion trap is the experimental few body problem. Present day ability to load a controlled number of atoms/molecules/ions in the lowest energy states of traps allows attempts at quantitative studies with these very challenging systems. Linear Paul traps have long supported ion crystals which order in shells about the trap axis. The 3-D ion trap would fill in a 3 dimensionally symmetric fashion~\cite{Alt98,Blo00,Sch00}. Problems such as minimum energy configurations, phonon mode excitations and crystal relaxation, rotations, effect of trap symmetry and its changes can be probed. The symmetry of the present trap can be dramatically altered from a totally symmetric to completely asymmetric configuration (i.e.  $\nu_{m;x}=\nu_{m;y}=\nu_{m;z}\leftrightarrow\nu_{m;x}\neq\nu_{m;y}\neq\nu_{m;z}$, with small differences in electrode voltages. In the highly asymmetric configuration the system can be driven into chaotic behavior as evident from single, trapped ion trajectory simulations, which exhibit a dense power spectrum. The ability to catch the ion signal in fluorescence makes trapped ions the natural choice for mapping both the equilibrium configurations and the dynamics of a system of a few, bound particles. Such studies with trapped ions are relevant across disciplines in physics.

In conclusion, the concept of ion trap described here is versatile and suitable for many different experiments over and above~\cite{Den05} those discussed here. The trap operating parameters determined here are easily scaled for different $Q/M$ and trap dimensions. Its primary strengths lie in the ease with which multiple, identical, ion traps can be realised and the ability to adapt the trap symmetry to a specific problem without additional experimental complexity.

\end{document}